\documentclass[twocolumn,a4paper]{article}
\usepackage{icqproc,epsf,amssymb,
epsfig}

\def\ts2f{${\cal T}^{*(2F)}$}
\def\t2f{${\cal T}^{(2F)}$}
\def\tsa4{${\cal T}^{*(A_4)}$}
\def\tms{${\cal T}^{(MS)}$}

\def\zfo{${{\mbox{$\bigcirc$}}\!\!\!\!\!\!\:{\mbox{2}}\,}$}

\def\ba{\begin{eqnarray}}
\def\ea{\end{eqnarray}}
\def\lb{\label}

\def\be{\begin{equation}}
\def\ee{\end{equation}}

\def\s{\sigma}
\def\a{\alpha}

\def\t{\tau}

\begin{document}
\thispagestyle{empty}

\twocolumn[
\vspace*{30mm}
\begin{LARGE}
\begin{center}
On Inflation Rules for Mosseri--Sadoc Tilings
\end{center}
\end{LARGE}
\begin{large}
\begin{center}
Zorka Papadopolos$^{\#}$,
Oleg Ogievetsky$^{\#\#}$
\end{center}
\end{large}
\begin{footnotesize}
\begin{it}
\begin{center}
$^{\#}$Institut f\"ur Theoretische Physik, Universit\"at
Magdeburg, PSF\ 4120, 39016\ Magdeburg, Germany\\
$^{\#\#}$Center of Theoretical Physics, Luminy,
13288 Marseille, France
\end{center}
\end{it}
\end{footnotesize}
\begin{footnotesize}
\begin{center}
Received 31 August 1999
\end{center}
\end{footnotesize}
\vspace{4ex}
\begin{small}
\hrule\vspace{3ex}
\begin{minipage}{\textwidth}
{\bf Abstract}\vspace{2ex}\\
\hp
We give the inflation rules for the decorated
Mosseri--Sadoc
tiles in the  projection class of tilings \tms.
Dehn invariants related to the stone inflation of the
Mosseri--Sadoc tiles provide eigenvectors of the
inflation matrix with eigenvalues equal to
$\t = \frac{1+\sqrt{5}}{2}$ and $(-\tau^{-1})$.
\vspace{2.5ex}
\end{minipage}\vspace{3ex}
\hrule
\end{small}\vspace{6ex}
]


\section{Introduction}

\hp

Kramer has introduced the icosahedrally symmetric tiling
of the (3dimensional) space by seven (proto)tiles~\cite{K}.
Sadoc and Mosseri have rebuilt these prototiles and reduced
their number to four: $z$, $h$, $s$ and $a$.
But the inflation class (inflation specie~\cite{D}) of the 
tilings of the space by the
four Mosseri--Sadoc prototiles has lost the icosahedral
symmetry~\cite{MS}. In Ref.~\cite{KP} a projection class
(projection specie~\cite{D}) of the tilings \tms \  by the 
Mosseri--Sadoc prototiles has been locally derived from the
canonical icosahedrally projected (from the lattice $D_6$)
local isomorphism class of the tilings \ts2f,~\cite{KPZ,P}.

In this paper we derive the inflation rules
for the decorated prototiles $z$, $r$, $m$, $s$ and $a$
of the projection class \tms. These rules enforce the
inflation rules for the {\em decorated} Mosseri--Sadoc
tiles $z$, $h$, $s$ and $a$ in agreement with those
suggested by Sadoc and Mosseri~\cite{MS}, up to the decoration.
Using the Dehn invariants~\cite{De}
for the prototiles (apart from the volumes of the prototiles)
we reproduce algebraically the inflation matrix for the
prototiles.

\section{Inflation of Decorated Mosseri--Sadoc Tiles}

Mosseri and Sadoc have given the inflation rules for their
$z$, $h$, $s$ and $a$ tiles~\cite{MS}. 
These rules were for the
Steininflation~\cite{D} of the tiles.
By definition, an inflation is a Steininflation if the 
inflated tiles are composed of the whole original tiles.

The inflation factor for the Mosseri--Sadoc tiles is
$\t=\frac{1+\sqrt{5}}{2}$.
The inflation matrix of the Steininflation of the tiles is
clearly the matrix with integer coefficients.
It has been given by Sadoc and Mosseri~\cite{MS}
\begin{eqnarray}
M=\left( \begin{array}{cccc}
    1&1&1&1\\
    2&1&2&2\\
    1&1&1&2\\
    0&0&1&2
\end{array} \right)
\ ,\lb{mmosa}
\end{eqnarray}
in the following ordering of the tiles: $z$, $h$, $s$ and $a$.

In the case of the Mosseri--Sadoc tiles, the Steininflation
is breaking the symmetry of the tiles.
The authors of Ref.~\cite{MS} haven't given a decoration of the
tiles which would  take care about the symmetry breaking and
{\em uniquely} define the inflation--deflation procedure
at {\em every step}.

%
%
\begin{figure}[]
\centerline{
}
\vfill
\centerline{
}
\vfill
\centerline{
}
\vfill
\centerline{
}
\vfill
\centerline{
}

\caption{
Decorated tiles $a$, $m$, $r$, $z$ and $s$ of the tilings
\tms \
obtained from the decorated eight tiles
$A^*$, $B^*$, $C^*$, $D^*$,  $F^{*r}$, $F^{*b}$,
$G^{*r}$ and $G^{*b}$  of the  tilings \ts2f.
The arrows which decorate the eight \ts2f--tiles
are along their edges, the arrows which decorate the
(composite) tiles of the projection species \tms \
are drawn on certain distance
in order to be distinguishable from the previous.
The ``white" arrow (by the tile $a$) is 
marking the edge $\tau^2$\zfo, the ``long" edge
in the $\tau$\ts2f--class of tilings.
}
\end{figure}

In~\cite{KP} it has been shown that the projection class
of the locally isomorphic tilings \ts2f,~\cite{KPZ}
(the tilings of the 3dimensional space by the six
tetrahedra with all edges parallel to the 2fold 
symmetry axis of an icosahedron, of two lengths,
the standard one denoted by \zfo, the ``short" edge, and
$\tau $ \zfo, the ``long" edge,~\cite{KPZ}) by the 
``golden" tetrahedra~\cite{OP}
can be locally transformed into the tilings \tms,
\ts2f $\longrightarrow $ \tms.
The class \tms \ of locally isomorphic tilings by 
Mosseri--Sadoc tiles
has been defined by the projection~\cite{KP}.
The important property is that the minimal packages of
the six golden tetrahedra in \ts2f,
such that their equilateral faces (orthogonal to the 
directions of the 3fold symmetry axis of an icosahedron) 
are covered, led to five tiles
$a$, $s$, $z$, $r$ and $m$~\cite{KP}.

In Fig. 1 we show how the {\em decorated}
prototiles~\cite{PHK}
$A^*$, $B^*$, $C^*$, $D^*$, $F^{*b}$, $F^{*r}$,
$G^{*b}$ and $G^{*r}$
of the projection class of tilings
\ts2f \ are locally transformed into the {\em decorated}
composite tiles denoted by $a$, $m$, $r$, $z$ and $s$.

%
%
\begin{figure}[]
\centerline{
}

\caption{ The tiles $r$ and $m$ appear in \tms \ always
as a union, $h = r\bigcup m$. The decoration of the tile
$h$ is determined by the decoration of the tiles
$r$ and $m$.}
\end{figure}

Moreover, the tiles $r$ and $m$ appear always as a union
$r\bigcup m$, that is, the tile $h$ of Sadoc and Mosseri with
three mirror symmetries~\cite{KP}. In Fig. 2 we show how the
decorated tiles $r$ and $m$ are locally transformed into a
decorated tile $h$. Hence, the projection class of tilings
\tms\ has four decorated prototiles  $a$, $h$, $z$ and $s$.
The decoration is such that it breaks all the symmetries of the
tiles.

From the inflation rules of the \ts2f \ tiles
$A^*$, $B^*$, $C^*$, $D^*$, $F^{*b}$, $F^{*r}$,
$G^{*b}$ and $G^{*r}$~\cite{PHK} we derive the inflation
rules for the five composite decorated prototiles
$a$, $m$, $r$, $z$ and $s$ and present them in Figs. 3--7.
Using the rule of the composition of the decorated tile $h$
given in the Fig. 2 one easily concludes how the decorated tile
$h$ inflates. The derived inflation rules of the four prototiles
$a$, $h$, $z$ and $s$ in the projection species \tms \ are,
up to a decoration, the same as those proposed by Sadoc and
Mosseri~\cite{MS}. Hence, we identify the inflation~\cite{MS}
and the projection species~\cite{KP} and denote them by the
same symbol, \tms.

\section{Dehn Invariants and Steininflation of
Mosseri--Sadoc Tiles}

The Dehn invariant of a polyhedron $P$~\cite{De} takes
values in a ring ${\bf R}\otimes {\bf R}_\pi$ where
${\bf R}_\pi$ is the additive
group of residues of real numbers modulo $\pi$; the tensor
product is over ${\bf Z}$, the ring of rational integers.
Denote by $l_i$ the lengths of edges of $P$.
Denote by $\a_i$  the corresponding lateral angles and by
$\bar{\a_i}$ -- the classes of $\a_i$ modulo $\pi$.
The Dehn invariant, ${\cal D}(P)$, of $P$ is equal to
\be {\cal D}(P)=\sum l_i\otimes \bar{\a_i}
\ ,\lb{di}\ee
with the sum over all edges of $P$.
The important property of Dehn invariants is their
additivity: if a polyhedron is cut in pieces, the Dehn 
invariant is the sum of the Dehn invariants of pieces.
This property implies that the vector of Dehn invariants 
of tiles is an eigenvector of the inflation matrix 
(\ref{mmosa}) with an eigenvalue $\tau$.
%
%
\begin{figure}
\centerline{
}
\caption{Inflation rule of the decorated tile $a$:
$\tau a = a \bigcup s \bigcup a$. The ``white" arrow
marks the edge $\tau^2$\zfo, the ``long" edge
in the  $\tau$\ts2f--class of tilings.}
\end{figure}
%
%
\begin{figure}
\centerline{
}
\caption{Inflation rule of the decorated tile $r$:
$\tau r = z \bigcup s \bigcup m \bigcup r$. }
\end{figure}
%
%
\begin{figure}
\centerline{
}
\caption{Inflation rule of the decorated tile $z$:
$\tau z = \tau r \bigcup a $. The white arrows are marking the
``short" and ``long" edges
in the $\tau$\ts2f--class of tilings.}
\end{figure}
%
%
\begin{figure}
\centerline{
}
\caption{Inflation rule of the decorated tile $s$:
$\tau s = \tau z \bigcup a $. The white arrow is marking the
``long" edge
in the  $\tau$\ts2f--class of tilings. }
\end{figure}
%
%
\begin{figure}
\centerline{
}
\caption{Inflation rule of the decorated tile $m$:
$\tau m = a \bigcup s \bigcup z \bigcup a $.
The white arrow is marking the ``long" edge
in the $\tau$\ts2f--class of tilings. }
\end{figure}

The vector of Dehn invariants for the Mosseri--Sadoc tiles is:
\be \vec{d}_{MS}={\cal D}
     \left( \begin{array}{c}z\\h\\s\\a\end{array}\right)
     =-5 \left( \begin{array}{c}
       \t\\ 2\\ \t-1\\-\t\end{array}\right)\otimes\bar{\a}\ ,
\lb{dms}\ee

\ba \cos \alpha &=&\frac{\t}{\t +2}=\frac{1}{\sqrt{5}}\ .
\ea
Thus, the space of Dehn invariants of the Mosseri--Sadoc tiles
is one--dimensional, there is only one independent 
lateral angle.
For the Dehn invariants applied to the inflation, see
Ref.~\cite{OP}


For the vector of volumes of the Mosseri--Sadoc tiles one
obtains
\be \vec{v}_{MS}={\rm Vol}
     \left( \begin{array}{c}z\\h\\s\\a\end{array}\right)
     =\frac{1}{12}\left( \begin{array}{c}
       4\t+2\\ 6\t+4\\ 4\t+3\\ 2\t+1\end{array}\right)
\ .\lb{vms}\ee

We show that the inflation matrix for
Mosseri--Sadoc tiles $z$, $h$, $s$ and $a$
can be uniquely reconstructed
from the Dehn invariants (and the volumes).

Denote the inflation matrix by $M_{MS}$.

The vectors $\vec{d}_{MS}$ and $\vec{v}_{MS}$
are eigenvectors of the inflation matrix, with the
eigenvalues $\t$ and $\t^3$
correspondingly (the eigenvalue is equal
to the inflation factor to the power which is the dimension
of the corresponding invariant).

Explicitely,
\be M_{MS}\left(
    \begin{array}{c}4\t+2\\6\t+4\\4\t+3\\2\t+1\end{array}\right)
   =\left(
    \begin{array}{c}16\t+10\\26\t+16\\18\t+11\\8\t+5
    \end{array}\right)
\lb{eigv}\ee
for the vector of volumes and
\be M_{MS}\left(\begin{array}{c}\t\\2\\ \t-1\\-\t
\end{array}\right)
   =\left(\begin{array}{c}\t+1\\2\t\\1\\-\t-1
\end{array}\right)\lb{eigd}\ee
for the the vector of Dehn invariants.

Assume that the inflation matrix is rational. 
Then, decomposing eqs. (\ref{eigv}) and (\ref{eigd})
in powers of $\tau$, we obtain  
four vector equations for $M_{MS}$ or a matrix equation
\be \begin{array}{rc}  
M_{MS}\left(\begin{array}{cccc}
     4&2&1&0\\
     6&4&0&2\\
     4&3&1&-1\\
     2&1&-1&0\end{array}\right)&\ \\[.7cm]
   =\left(\begin{array}{cccc}
     16&10&1&1\\
     26&16&2&0\\
    18&11&0&1\\
    8&5&-1&-1\end{array}\right)&\ .
\end{array}
\lb{mms}
\ee
The solution of this equation is unique and we find the
matrix (\ref{mmosa}).


\vskip .3cm

\section*{Acknowledgements}

The work of Z. Papadopolos was supported by the Deutsche 
Forschungsgemeinschaft.
Z. Papadopolos is grateful for the hospitality to the center
of the Theoretical Physics in Marseille, where this work has
been started.
The work of O. Ogievetsky was supported by the Procope 
grant 99082.
We also thank the Geometry--Center at the University of 
Minnesota for making Geomview freely available.


\hp

\begin{footnotesize}
\begin{frenchspacing}

\end{frenchspacing}
\end{footnotesize}

\end{document}